\documentclass[useAMS,usenatbib]{mn2e}
\pdfoutput=1
\usepackage{latexsym}
\usepackage{amsmath}
\usepackage{amssymb}
\usepackage{fnpos}
\usepackage{scrextend}

\usepackage{tabularx}
\usepackage{xspace}
\usepackage{enumerate}
\usepackage{graphicx}
\usepackage{caption}
\usepackage{gensymb}
\usepackage{subcaption}
\usepackage{longtable}
\usepackage{lscape}

\newcommand{\eal}[2]{\ifmmode{\mathrm{#1\,#2}}\else{#1\textsc{$\,$\lowercase{#2}}}\fi\xspace}
\newcommand{\feal}[2]{\ifmmode{\mathrm{#1\,#2}}\else{[#1\textsc{$\,$\lowercase{#2}}]}\fi\xspace}

\title[V5852 Sgr: An Unusual Nova]{V5852 Sgr: An Unusual Nova Possibly Associated with the Sagittarius Stream}

\author[Aydi et al.]{E. Aydi$^{1,2}$\thanks{E-mail: eaydi@saao.ac.za}, P. Mr\'oz$^{3}$, P. A. Whitelock$^{1,2}$, S. Mohamed$^{1}$, \L{}. Wyrzykowski$^{3}$,A. Udalski$^{3}$,  \newauthor   P. Vaisanen$^{1,4}$, T. Nagayama$^{5}$, M. Dominik$^{6}$,   A. Scholz$^{6}$,  H. Onozato$^{7}$,  R. E. Williams$^{8}$,\newauthor S. T. Hodgkin$^{9}$, S. Nishiyama$^{10}$, M. Yamagishi$^{11}$,  A. M. S. Smith$^{12,13}$, T. Ryu$^{14,15}$, \newauthor A. Iwamatsu$^{10}$, and I. Kawamata$^{10}$\\ 
$^{1}$South African Astronomical Observatory, P.O. Box 9, 7935 Observatory, South Africa\\
$^{2}$Astronomy Department, University of Cape Town, 7701 Rondebosch, South Africa\\
$^{3}$Warsaw University Observatory, Al. Ujazdowskie 4, 00-478 Warszawa, Poland\\
$^{4}$Southern African Large Telescope, P.O. Box 9, 7935 Observatory, South Africa\\
$^{5}$Department of Physics and Astronomy, Graduate School of Science and Engineering, Kagoshima University, 1-21-35 Korimoto, \\\,\,\,\,\,Kagoshima 890-0065, Japan\\
$^{6}$SUPA, School of Physics \& Astronomy, University of St Andrews, North Haugh, St Andrews, KY169SS, UK.\\
$^{7}$Astronomical Institute, Graduate School of Science, Tohoku University 6-3 Aramaki Aoba, Aoba ku, Sendai, Miyagi 980-8578, Japan\\
$^{8}$Space Telescope Science Institute, 3700 San Martin Drive, Baltimore, MD 21218, USA\\
$^{9}$Institute of Astronomy, University of Cambridge, Madingley Rise, Cambridge, CB3 0HA, UK\\
$^{10}$Miyagi University of Education, Aoba-ku, Sendai, Miyagi 980-0845, Japan\\
$^{11}$Graduate School of Science, Nagoya University, Furo-cho, Chikusa-ku, Nagoya 464-8602, Japan\\
$^{12}$N. Copernicus Astronomical Centre, Polish Academy of Sciences, Bartycka 18, 00-716, Warsaw, Poland\\
$^{13}$Institute of Planetary Research, German Aerospace Center, Rutherfordstrasse 2, 12489 Berlin, Germany\\
$^{14}$SOKENDAI, The Graduate University for Advanced Studies, 2-21-1 Osawa, Mitaka, Tokyo 181-8588, Japan\\
$^{15}$National Astronomical Observatory of Japan, 2-21-1 Osawa, Mitaka, Tokyo 181-8588, Japan\\
}

\begin{document}

\date{Accepted: 08 June, 2016. Received ***; in original form 2015 December 30}
\pagerange{\pageref{firstpage}--\pageref{lastpage}} \pubyear{2015}
\maketitle

\label{firstpage}
\begin{abstract}
We report spectroscopic  and photometric follow-up of the peculiar nova V5852~Sgr (discovered as OGLE-2015-NOVA-01), which exhibits a combination of features from different nova classes. The photometry shows a flat-topped light curve with quasi-periodic oscillations, then a smooth decline followed by two fainter recoveries in brightness. Spectroscopy with the Southern African Large Telescope shows first a classical nova with an \eal{Fe}{II} or \eal{Fe}{II} b spectral type. In the later spectrum, broad emissions from helium, nitrogen and oxygen are prominent and the iron has faded which could be an indication to the start of the nebular phase. The line widths suggest ejection velocities around $1000\,{\rm km\,s^{-1}}$. The nova is in the direction of the Galactic bulge and is heavily reddened by an uncertain amount. 
The $V$ magnitude 16 days after maximum enables a distance to be estimated and this suggests that the nova may be in the extreme trailing stream of the Sagittarius dwarf spheroidal galaxy. If so it is the first nova to be detected from that, or from any dwarf spheroidal galaxy. Given the uncertainty of the method and the unusual light curve we cannot rule out the possibility that it is in the bulge or even the Galactic disk behind the bulge.
\end{abstract}


\section{Introduction}
\label{Intro}

Classical novae (CNe) form a subclass of the cataclysmic variable stars. They are ``close binary systems consisting of a white dwarf (WD) and a Roche-lobe filling companion'' \citep{Bode_etal_2008}. Matter accreted by the WD from the secondary accumulates and is compressed on the WD surface. When the critical temperature and density are reached a thermonuclear runaway (TNR) is triggered, causing a nova eruption \citep{Prialnik_1986}. Typical ejecta velocities and masses are $\sim$ 1000 $\mathrm{km\,s^{-1}}$  and between  $10^{-5}$ and $10^{-4} \,\mathrm{M_{\odot}}$, respectively \citep{Payne-Gaposchkin_1957,Gallaher_etal_1978}.

Nova light curves are classified on the basis of their decline rate in various ways, but often in terms of time in days ($t_2$) for the decline by 2 mag from maximum light \citep{Payne-Gaposchkin_1964}. More recently, \citet{Strope_etal_2010} published an Atlas of light curves of 93 novae based on visual estimates collected by AAVSO and they proposed a classification scheme based on the light curve shape and rate of decline. They defined seven classes: S  (smooth), P (plateau), D (dust dip), C (cusped secondary maximum), O (oscillations), F (flat-topped), and J (jitters). It is important to recognize that this classification is based on visual light curves and given that the colours of novae typically change during their evolution we should not expect other bands to track the visual curves with any degree of precision.  Nevertheless, this nomenclature has been used by others for $I$-band light curves of novae (e.g.\citet{Mroz_etal_2015} where they classified OGLE $I$-band light curves of 39 novae in the Galactic bulge using the same scheme) and we therefore refer to it here.

In addition to the photometric classification various approaches have been taken to classify the spectral evolution of novae \citep{McLaughlin_1944,Payne-Gaposchkin_1957,Williams_2012}. Post-outburst spectra are divided into two spectroscopic classes: the \eal{Fe}{II} type novae which show numerous narrow \eal{Fe}{II} emission lines with P Cygni profiles and He/N type novae which show broad He, N and H emission lines. A third class was also introduced, known as hybrid novae. These show a transition between the two types (\eal{Fe}{II} and He/N) or simultaneous emission lines of both types \citep{Williams_1992}.
From a study of 22 Galactic novae, \citet{Della_Valle_etal_1998} established that novae showing He/N spectroscopic features are faster, and have brighter maxima, compared to those from the \eal{Fe}{II} class. The precise mechanism responsible for the formation of the different spectral features after the explosion is poorly understood, but it is thought that the WD mass and the circumstellar environment play a crucial role in this matter \citep{Shafter_etal_2011}.

In this paper, we report the discovery of an unusual nova, V5852~Sgr, that may be a member of the trailing stream of the Sagittarius Dwarf Spheroidal Galaxy (henceforth the Sagittarius stream). At about 20\,kpc the Sagittarius Dwarf is our nearest neighbour and is being tidally disrupted by its interaction with the Milky Way. It was discovered by \citet{Ibata_etal_1995} via the distinctive velocities of its stars. It has since been shown to be stretched into a stream that loops around the Milky Way, with an estimated total mass of about $5 \times 10^8\ M_{\odot}$ \citep{Majewski_etal_2003}, making it the largest dwarf spheroidal galaxy in the Local Group. 

The nova shows combined features from different novae spectro-photometric classes and  presents an interesting case to study. In Section~\ref{photo} we present the discovery and the photometric (optical and IR) follow-up of the object.  SALT (Southern African Large Telescope) spectroscopic optical data (between 4500 \AA\ and 7800 \AA) are detailed in Section~\ref{Spectro}. We present in Section~\ref{Concl}. a discussion aiming to classify the nova, and establish its location and our conclusions.

\section{Discovery and Photometric data}
\label{photo}
\subsection{Discovery}
\begin{figure}
\centering
\begin{subfigure}{.5\textwidth}
  \centering
  \includegraphics[width=.7\linewidth]{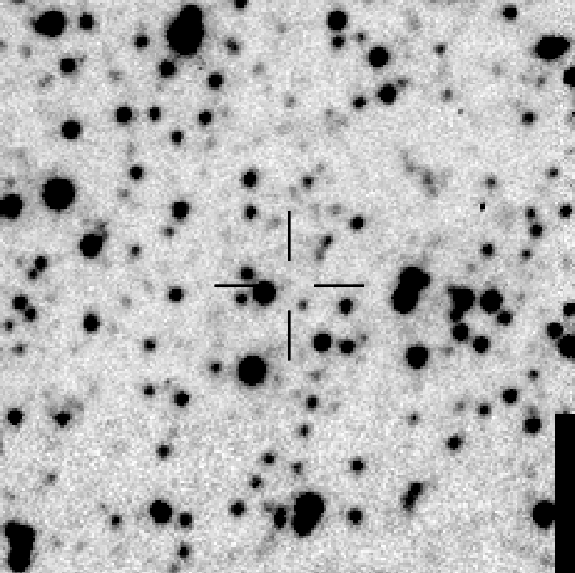}
  \caption{Finding chart before the eruption.}
  \label{fig:sub1}
\end{subfigure}
\begin{subfigure}{.5\textwidth}
  \centering
  \includegraphics[width=.7\linewidth]{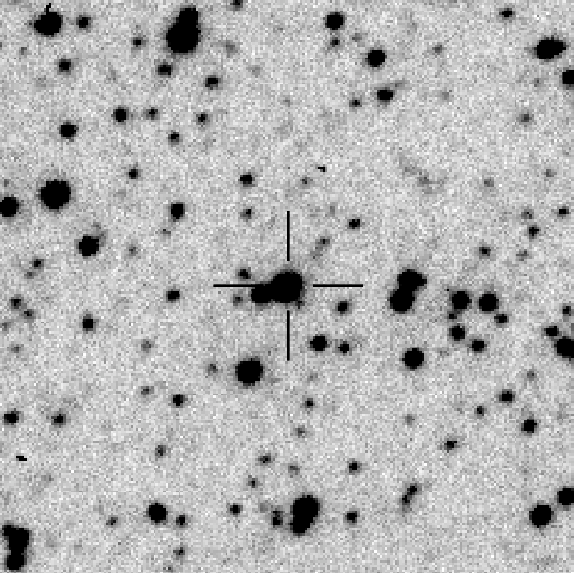}
  \caption{Finding chart during the eruption (5$^{\rm th}$ of March 2015).}
  \label{fig:sub2}
\end{subfigure}
\caption{Finding charts for V5852~Sgr before and during the eruption in the $I$ band. (North is up and East is to the left, image size $1' \times 1'$).}
\label{fig1first}
\end{figure}

V5852~Sgr  was announced as a classical nova candidate and given the name OGLE-2015-NOVA-01, on 2015 March 5 by \citet{ATel_7179} based on observations from the Optical Gravitational Lensing Experiment (OGLE) survey. The transient was detected by the Early Warning System (EWS), which has been designed for the detection of microlensing events. Fig.~\ref{fig1first} presents charts of the nova before and during the eruption. From the time of the discovery, however, it was clear that the light curve (Fig.~\ref{Fig:1}) does not resemble that of a microlensing event. The star is located in the direction of the Galactic bulge at equatorial coordinates of $(\alpha, \delta)_{\rm J2000.0}$ = (17h:48m:12.78s, --32$^{\circ}$:35':13''.44)  and Galactic coordinates of ($l, b$) = (357$^{\circ}$.16, --2$^{\circ}$.36). 
\subsection{Observation and data reduction}

The source has been observed by the OGLE variability survey since 2010 with a typical cadence of 1--2 d. This survey uses the 1.3-m Warsaw Telescope located at Las Campanas Observatory, Chile, operated by the Carnegie Institution for Science. Observations were taken through $V$- and $I$-band filters, closely resembling those of the standard Johnson-Cousins system. The photometry was carried out with the Difference Image Analysis (DIA) algorithm \citep{Alard_etal_1998,Wozniak_2000}. Details of reductions and calibrations can be found in \citet{Udalski_etal_2015}. The light curve of the 2015 eruption data is shown in Fig.~\ref{Fig:1}.  In Tables~\ref{table2} and ~\ref{iband} we present OGLE $V$- and $I$-band photometry, respectively.

Multicolor $gri$ photometry was obtained using the LCOGT 1-m robotic telescope at the South African Astronomical Observatory (SAAO), Sutherland, South Africa on 2015 March 19 and April 22-23. The data are presented in Table~\ref{table1} and included in Fig.~\ref{Fig:1}. The photometry was carried out with DoPHOT \citep{Schechter_etal_1993} and calibrated with the aid of the AAVSO Photometric All-Sky Survey (APASS \footnote{https://www.aavso.org/apass}) \citep{Henden_etal_2009}.

The source was observed by the \textit{Swift} satellite \citep{Gehrels_etal_2004}  on 2015 April 11, 15, and 20 with a total exposure time of 2.61~ks. No X-ray source was detected with the XRT (X-Ray Telescope; \citealt{Burrows_etal_2005}) at the position of the nova, which gives $3\sigma$ upper limits of 0.022, 0.024, and 0.008 cts/s, respectively, in the 0.3-10 keV energy range. Upper limits were estimated with the {\sc sosta} tool, which is a part of the XIMAGE package (version 4.5.1).

We also analyzed images taken with the UVOT (Ultraviolet/Optical Telescope; \citealt{Roming_etal_2005}) on-board {\it Swift} and estimated the brightness of the source with the {\sc uvotsource} tool, which performs aperture photometry. We detected a faint counterpart in the $U$-band ($21.03 \pm 0.58$ mag), while the source was invisible in the $UVW1$ filter (which peaks at 260 nm) images. The UVOT observations are summarized in Table~\ref{table1}.

We obtained infrared photometry with the SIRIUS camera \citep{Nagayam_etal_2003} on the 1.4-m Japanese-South African InfraRed Survey Facility (IRSF) at SAAO Sutherland. This has a $7' \times 7'$ field of view, $0''.45$  pixels and provides simultaneous $JHK_S$ photometry. Exposures of 125\,s were achieved by combining 25 dithered 5\,s exposures. Photometry on the 2MASS system was performed relative to the nearby stars 2MASS17481467--3235040 and 2MASS17481308--3235223. 
The rms of the instrumental magnitudes of these reference stars was only $\sigma_J = 0.010, \ \sigma_H = 0.008, \ \sigma_K = 0.007$ mag, demonstrating that they do not vary.

As can be seen in Fig.~\ref{fig1first}, the field is crowded, in particular the nova is only $2''.8$ from 2MASS17481299--3235142 ($J=12.9 \ H=11.4 \ K_S=10.8$). So to do aperture photometry the stars in the immediate vicinity of the nova, and of the two reference stars, were subtracted using {\sc DAOPHOT} in {\sc  IRAF} before the photometry was performed. The 2MASS magnitudes of the reference stars were then used to derive the zero points and an average taken. Photometry was done using apertures with radii of 5, 7 and 10 pixels. For  measurements of $J,H<14$ and $K_S<13.5$ the choice of aperture makes no significant difference to the result. For much fainter magnitudes the results from the 5 and 10 pixel apertures can differ by up to 0.2 mag.  The measurements listed in Table~\ref{table3} were made with a 7 pixel aperture.

\subsection{Light curve}
\label{lc}
The eruption started between 2015 February 25 (HJD 2457078.88) and 27 (HJD2457080.90), when the source brightened by 2 mag from $I=19.7$ to $I=17.6$; by March 3 (HJD 2457084.8) it had reached $I=14.4$. This was followed by a short pre-maximum halt and then a nearly linear rise by 1.5 mag in 14 days. The source peaked on 2015 April 3 (HJD=2457115.87)  at $I=12.7$ (although the maximum could have taken place ten days earlier, when no observations were collected because of heavy storms over Las Campanas). Between 2015 March 20 and May 1, the source showed a long, flat peak, similar to F class light curves, with additional semi-regular oscillations on a time-scale of 12\,d and amplitude of ~0.8 mag, reminiscent of the early stages of D class (e.g. V992 Sco) or  J class light curves \citep{Strope_etal_2010}. At this phase the light curve also resembled the early stages of the outburst of the red nova V838 Mon (\citealt{Munari_etal_2002} fig.~5) and the 2011 outburst of the recurrent nova T Pyx (\citealt{Surina_etal_2014}  fig.~1). 

\begin{figure*}
\begin{center}
\includegraphics[width=160mm]{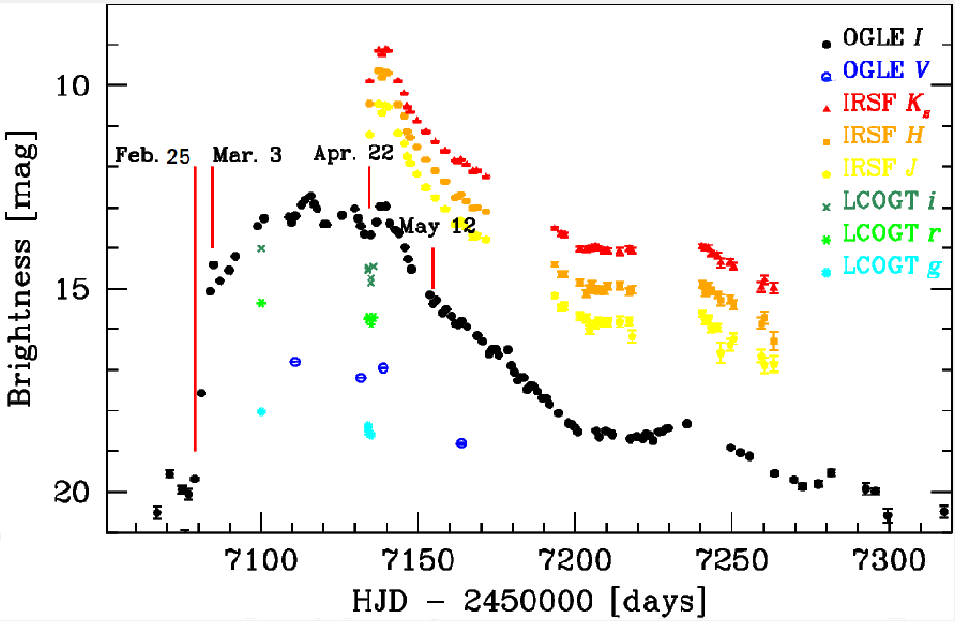}
\caption{The photometric data from OGLE, LCOGT, and IRSF as a function of Heliocentric Julian Date (HJD),  colour coded as indicated.
The first two red bars indicate the dates when the eruption starts. The SALT spectra were obtained on April 22 and May 12.}
\label{Fig:1}
\end{center}
\end{figure*}

\begin{figure}
\begin{center}
\includegraphics[width=70mm]{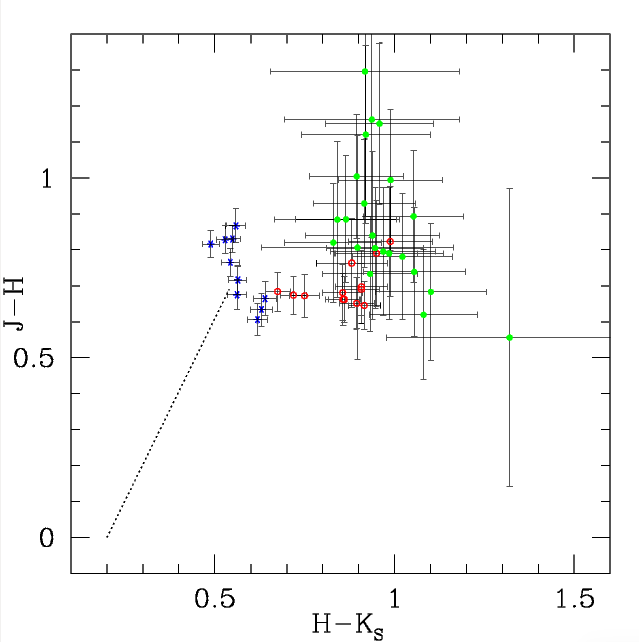}
\caption{Evolution of the $JHK_S$ colours; asterisks, open circles and closed circles represent observations before HJD 2457150, between HJD 2457150 and 2457200 and after HJD 2457200, respectively. The dotted vector shows the effect of correcting for reddening assuming $E(J-H)=0.69$ and $E(H-K_S)=0.34$ (see section~\ref{red})}
\label{JHK}
\end{center}
\end{figure}

Infrared $JHK_S$ (Table~\ref{table3}) measurements were only initiated when it was clear that the $I$ light curve development was unusual and, as can be seen in Fig.~\ref{Fig:1}, the $JHK_S$-curves, more or less follow the fall of the $I$ curve.
Fig.~\ref{JHK} shows the development of the infrared colours. The reddening vector indicates that  $(J-H)_0$ is close to zero or slightly negative and that it is remarkably constant, while $(H-K_S)_0$ gets slightly redder with time. A negative value for $(J-H)_0$ has been seen in other novae \citep{Whitelock_etal_1984} and is attributed to very strong emission lines in the $J$ band (the SAAO $J$ filter used by Whitelock et al. is more sensitive to the He{\sc I} 1.083 $\mu$m line than is the IRSF $J$ filter and therefore their $J-H$ is very negative when this line is strong). Given that novae spectra are dominated by emission lines it is not possible to properly transform photometry taken through different filters, so detailed comparisons of nova colours are difficult.

It is obvious that the overall development of the light curve (Fig.~\ref{Fig:1}) is quite unlike any of the standard nova classes as can be seen by comparing it with the OGLE light curves of bulge novae \citep{Mroz_etal_2015}. The decline at around HJD2457150 and the recovery around HJD2457230 could possibly have been the consequence of dust forming and gradually dispersing. However, Fig.~\ref{JHK} shows no indication of the very red colours associated with significant quantities of dust. The colours of dust-forming novae are illustrated in fig.~4 of \citet{Whitelock_etal_1984} for several other novae.  

We measured decline times in the $I$ band over 2 and 3 mags of $t_2 = 34 \pm 2$\,days and $t_3 = 45 \pm 2$\,days, respectively. So this nova was moderately fast to average according to the criteria in \citet{Warner_2008} or \citet{Bode_etal_2008}, although given the unusual characteristics of the light curve and the fact that we only have decay rates at $I$, and not $V$, it is not entirely clear how this should be interpreted. 

In the pre-eruption images we detected a possible very faint progenitor $I_{\rm pre}=20.77$, which is invisible in the $V$-frames (as expected given the reddening discussed in section~\ref{red}). We did not find any significant periodic variability before the eruption. 

In summary, our photometric data show a moderately fast outburst, followed by a flat-topped light curve with oscillations and  a smooth decline followed by two faint recoveries in the brightness (between HJD 2457220 and 2457270 and between 2457280 and 2457300). A comparison with the light curve classes presented by \citet{Strope_etal_2010} shows that the flat-topped light curve of V5852 Sgr is similar to the F class light curves. In addition, the quasi-periodic oscillations during the peak in the intensity are similar to both J class and O class light curves. This combination of features indicate the unusual light curve development of V5852 Sgr. Note that a flat-topped light curve with oscillations is also seen for the 2011 light curve of the recurrent nova T Pyx. A detailed spectro-photometric study by \citet{Surina_etal_2014} of the T Pyx 2011 outburst shows a pre-maximum halt followed by a slow flat-topped outburst with quasi-periodic oscillations of a period $\sim$\,2 days.
\\
The remarkable features in the light curve justified our request for SALT spectra to establish if the spectral development was also unusual.

\begin{table}
\centering
\begin{tabular}{rrr}
\hline
HJD & $V$ & $\Delta V$ \\ 
-2457000 & \multicolumn{2}{c}{(mag)}\\
\hline
110.84656 & 16.799 & 0.005 \\
131.79511 & 17.184 & 0.006 \\
138.84101 & 16.924 & 0.005 \\ 
163.78319 & 18.945 & 0.017 \\ \hline
\end{tabular}
\caption{OGLE $V$-band photometry. The time-series photometry is available from the OGLE Internet Archive\protect\footnotemark.}
\label{table2}
\end{table}

\begin{table}
\centering
\begin{tabular}{rrr}
\hline
HJD & $I$ & $\Delta I$\\
-2457000 & \multicolumn{2}{c}{(mag)}\\
\hline
66.86475 & 20.507 & 0.153\\
70.89930 & 19.559 & 0.092\\
74.85976 & 19.941 & 0.089\\
75.83561 & 21.281 & 0.335\\
76.90152 & 20.055 & 0.135\\
78.87609 & 19.678 & 0.043\\
80.89831 & 17.569 & 0.010\\
83.80943 & 15.059 & 0.002\\
\hline
\end{tabular}
\caption{A sample of OGLE $I$ photometry. The rest of the data can be found on the electronic version and are available from the OGLE Internet Archive$^3$.}
\label{iband}
\end{table}

\footnotetext{ftp://ftp.astrouw.edu.pl/ogle/ogle4/NOVAE/BLG}

\begin{table}
\centering
\begin{tabular}{lrrr}
\hline
HJD & Filter & Exp. time  &  Magnitude  \\
-2457000 & & (s) &\\
\hline
LCOGT \\ \hline
100.60483 &	$g$ &	450 &	18.026 +/- 0.011 \\
100.61375 &	$r$ &	180 &	15.366 +/- 0.006 \\
100.61103 &	$i$ &	180 & 	13.982 +/- 0.004 \\
134.54530 &	$g$ &	200 &	18.552 +/- 0.054 \\
134.54815 &	$i$ &	100 &	14.545 +/- 0.017 \\ 
134.55012 &	$r$ &	100 &	15.751 +/- 0.022 \\
134.57024 &	$g$ &	200 &	18.405 +/- 0.033 \\
134.60703 &	$g$ &	200 &	18.468 +/- 0.043 \\
134.60988 &	$i$ &	100 &	14.513 +/- 0.029 \\
134.61164 &	$r$ &	100 &	15.749 +/- 0.028 \\
134.64962 &	$g$ &	200 &	18.365 +/- 0.065 \\
134.65258 &	$i$ &	100 &	14.499 +/- 0.028 \\
134.65433 &	$r$ &	100 &	15.712 +/- 0.033 \\
135.49551 &	$i$ &	100 &	14.858 +/- 0.035 \\
135.52847 &	$g$ &	200 &	18.598 +/- 0.040 \\
135.53115 &	$i$ &	100 &	14.739 +/- 0.049 \\
135.53302 &	$r$ &	100 &	15.870 +/- 0.024 \\
136.41910 &	$i$ &	100 &	14.462 +/- 0.053 \\
136.42073 &	$r$ &	100 &	15.712 +/- 0.056 \\ \hline
{\it Swift} \\ \hline
124.36    & $UVW1$ &    720 &   $> 20.07^{\rm a}$ \\
127.52    & $UVW1$ &    479 &   $> 19.85^{\rm a}$ \\
133.41    & $U$  &   1345 &    21.03 +/- 0.58       \\
\hline
\end{tabular}  \\
\medskip
$^{\rm a}$ $3\sigma$ limit
\caption{LCOGT and {\it Swift} photometry of V5852~Sgr.}
\label{table1}
\end{table}

\begin{table}
\begin{center}
\caption{Near-infrared photometry from IRSF at $JHK_S$ bands.}
\label{table3}
\begin{tabular}{ccccccc}
\hline
HJD  &  $ J$ &  $\Delta J$  & $ H$ &  $\Delta H$ & $K_S$ & $\Delta K_S$\\
-2457000 &\multicolumn{6}{c}{(mag)}\\
\hline
134.52332 &11.21 &0.02 &10.45 &0.01  &9.91 &0.01\\
137.49937 &10.45 &0.02  &9.64 &0.01  &9.15 &0.01\\
138.46439 &10.67 &0.03   &9.81 &0.02  &9.25 &0.01\\
139.55240 &10.49 &0.02  &9.67 &0.01  &9.14 &0.01\\
140.40741 & 10.52 &0.02&  9.69 &0.01 & 9.15 &0.01\\
143.61646 &11.19 &0.02 &10.47  &0.01  &9.91 &0.01\\
145.55649 &11.43 &0.03 &10.76  &0.01 &10.20 &0.01\\
146.50150 &11.75 &0.03 &11.14 &0.02 &10.52 &0.01\\
147.43051 &11.91 &0.03 &11.28 &0.02 &10.65 &0.01\\
149.66654 &12.18 &0.03  &11.51 &0.02 &10.87 &0.01\\
152.49157 &12.51 &0.03 &11.82 &0.02  &11.15 &0.01\\
155.45161 &12.77 &0.03 &12.10 &0.02  &11.38 &0.02\\
158.57664 &13.05 &0.04 &12.37 &0.02 &11.62 &0.02\\
161.62666 &13.41 &0.04 &12.77 &0.03 &11.85 &0.02\\
162.64767 &13.40 &0.04 &12.74 &0.03 &11.88 &0.02\\
163.60868 &13.34 &0.04 &12.68 &0.02 &11.82 &0.02\\
165.41369 &13.50 &0.04 &12.85 &0.03 &11.95 &0.02\\
167.57171 &13.72 &0.06 &13.03 &0.04 &12.12 &0.03\\
168.65172 &13.70 &0.04 &13.01 &0.03 &12.10 &0.02\\
171.65273 &13.79 &0.05 &13.11 &0.03 &12.25 &0.02\\
193.51274 &15.17 &0.07 &14.40 &0.05 &13.52 &0.04\\
195.61862 &15.46 &0.08 &14.64 &0.07 &13.65 &0.05\\
196.61061 &15.43 &0.08 &14.64 &0.07 &13.69 &0.05\\
201.53558 &15.68 &0.09  &14.86 &0.08&14.03 &0.06\\
203.59956 &15.74 &0.09 &15.12 &0.09 &14.04 &0.06\\
204.60055 &16.00 &0.11  &15.01 &0.09 &14.02 &0.06\\
205.31855 &15.89  &0.10 &14.89 &0.07 &13.99 &0.06\\
206.37154 &15.90 &0.10 &15.01 &0.08 &13.96 &0.06\\
207.34053 &15.81 &0.09 &15.03 &0.08 &14.01 &0.06\\
209.26551 &15.83 &0.09 &15.04 &0.09 &14.07 &0.06\\
210.41849 &15.82 &0.10 &14.94 &0.08 &14.07 &0.06\\
214.23445 &15.81 &0.13 &14.93 &0.09 &14.09 &0.09\\
217.27441 &15.81 &0.10 &15.07 &0.08 &14.01 &0.06\\
218.19040 &16.17 &0.14 &15.02 &0.08 &14.06 &0.06\\
240.39200 &15.62 &0.08 &14.89 &0.08 &13.96 &0.06\\
241.47998 &15.78 &0.10 &15.09 &0.09 &13.99 &0.06\\
242.31996 &15.75 &0.09 &14.94 &0.08 &13.99 &0.06\\
243.40194 &15.98 &0.10 &15.05 &0.08  &14.13 &0.06\\
245.38289 &15.95 &0.10 &15.16 &0.08 &14.18 &0.07\\
246.43287 &16.58 &0.27 &15.28 &0.14 &14.37 &0.13\\
249.44780 &16.38  &0.15 &15.26 &0.10  &14.34 &0.08\\
250.45178 &16.23 &0.13 &15.39 &0.10 &14.45 &0.08\\
259.29456 &16.66 &0.16 &15.86 &0.15 &14.96 &0.12\\
260.31153 &16.88 &0.19  &15.71 &0.14 &14.78 &0.10\\
263.28446 &16.85 &0.19 &16.29 &0.22 &14.97 &0.12\\
\hline
\end{tabular}
\end{center}
\end{table}

\subsection{Reddening and Distance}
\label{red}
Given its position towards the bulge, we expect the nova to be heavily reddened;
determining exact values poses a challenge.  At  $l=-2^{\rm o}.8$ and $b=-2^{\rm o}.4$ it is very close, but just outside the region around the Galactic centre ($|l| < 3^{\rm o}$, $|b| < 1^{\rm o}$) for which \citet{Nishiyama_etal_2009} determined an interstellar extinction law that is significantly different from that generally assumed \citep{Cardelli_etal_1989}. Recently, \citet{Nataf_etal_2015} have shown that large variations in the extinction ratios are common around the Galactic centre.

The VVV extinction calculator, which is based on \citet{Gonzalez_etal_2012}, indicates $E(J-K_S)=1.03$ corresponding to $A_K=0.71\pm0.12$ and $A_V=6.44$ mag on the Cardelli law or $A_K=0.54\pm0.12$ and $A_V=8.88$ mag on the Nishiyama law. As an alternative approach we measured the centroid of the red giant clump $(V-I)_{\rm clump} = 3.99 \pm 0.03$ on the color-magnitude diagram, using the OGLE data for stars in an area of $2'\times 2'$ around the nova. The intrinsic color of the Galactic bulge red clump stars is 1.06 (e.g., \citealt{Nataf_etal_2013}), so the color excess is $E(V-I)=2.93 \pm 0.03$ in this direction. Using the extinction relations for the Galactic bulge from \citet{Nataf_etal_2013}, we find $A_I = 3.60 \pm 0.17$ mag and $A_V = 6.53 \pm 0.18$ mag. Although this method estimates the extinction foreground of the bulge red clump stars, we can reasonably assume that most of the extinction is in the foreground disc and therefore included. This method should provide the best possible measure for the line-of-site to the nova and therefore in the following analysis we assume: $A_V=6.53$ and $A_I=3.60$. It is important to recognize that many of the conclusions of this paper are critically dependent on this assumption. 

The distances to novae are often estimated from their decline rates, noting that fast novae are 
more  luminous than slow ones. However, the individual
measurements have considerable uncertainties and the validity of the
so-called ``maximum magnitude versus rate of decline'' (MMRD) relations was 
recently questioned by \citet{Kasliwal_etal_2011}, \citet{Cao_etal_2012}, and \citet{Shara_etal_2016}. Furthermore, these relations have not been defined in the $I$-band and no empirical relation between the rate of decline in $I$ and $V$ has been derived (Munari private communication).

A relationship sometimes used to derive distances is based on the understanding that all novae have a similar $V$ mag 15 days after maximum: $M_V=-6\pm0.44$ \citep{Downes_etal_2000}. Coincidentally V5852~Sgr was measured at $V$ about 16 days after maximum (Table~\ref{table2}) with $V=17.18$ (making the reasonable assumption that maximum at $V$ and $I$ are close together in time), corresponding to $V_0=10.65$ and a distance ($(m-M)_0=16.65$ mag) in the range 17 to 27\, kpc.  Because of the unusual light curve  (section \ref{lc}) and the extension in the peak of brightness with quasi-periodic oscillations, this distance must be regarded as very uncertain.

There are several possible explanations of the large distance estimate.  Most obviously it suggests the nova is behind the bulge, rather than in it. At about 22\,kpc it would be about the right distance to be associated with the Sagittarius stream \citep{Ibata_etal_1995}.  Although it is a few degrees away from where the Sagittarius stream crosses the Galactic plane, it is within the area where the OGLE group found RR Lyr variables that are members of the Sagittarius stream  (see fig.~3 in \citet{Scozynski_etal_2014}). The nova is not coincident with the main density of OGLE variables (with distances from \citealt{Pietrukowicz_etal_2015}), but it is plausibly within the volume occupied by the trailing stream from the Sagittarius dwarf  which can be seen 
from \citet{Torreabla_etal_2015} when their fig. 16 is interpolated through the Galactic plane.
Furthermore, the radial velocity (see Section~\ref{specvel}) although uncertain is consistent with membership of the Sagittarius stream. 

We need to consider other possible explanations as alternatives to membership of the Sagittarius stream. First it could be in the Galactic disk behind the bulge; at 22\,kpc it is less than 1\,kpc above the plane. This is discussed further in  Section~\ref{specvel} in the context of the nova's radial velocity. Secondly, given its peculiar light curve this nova may not follow the usual decay-rate luminosity relations and therefore could be in the bulge. 
Thirdly, because of the peculiarity of the reddening law discussed above it is possible that the interstellar extinction at $V$ is higher than the $A_V=6.53$ used here and therefore that the nova is in the bulge and not sub-luminous. Although values as high as the $A_V = 8.88$ mag derived from the Nishiyama law can be ruled out, because in that case the red clump stars would not have been observable.

Some combination of greater reddening and slightly sub-luminous nova could also explain the observations. It has been suggested that novae in the bulge, which originate from low mass white dwarfs, are somewhat fainter than those in the disk (e.g. \citet{Della_Valle_etal_1998} and \citet{Della_Valle_2002}). There is also some evidence that relatively slow novae are more common in the bulge and fast novae in the disk \citep{Shafter_etal_2011}.

\section{Spectroscopic Data}
\label{Spectro}
\subsection{Observation and Data Reduction}
The nova was observed on 2015 April 22 and May 12, using the Robert Stobie Spectrograph (RSS; \citealt{Burgh_etal_2003}; \citealt{Kobulnicky_etal_2003}), mounted on the  Southern African Large Telescope (SALT) situated at the SAAO, Sutherland, South Africa. The observation on 2015 April 22 consists of two spectral ranges; [4500 $\mathrm{\AA}$ -- 5850 $\mathrm{\AA}$] and [5800 $\mathrm{\AA}$ -- 7000 $\mathrm{\AA}$]. The RSS long-slit mode was used with a $0.6''$ slit at a resolution of $R \sim 5000$ for both spectral ranges with exposure times of (4 $\times$ 100 s) and (2 $\times$ 150 s), respectively. The poor weather conditions and seeing resulted in a limited S/N.

The May 12 observation was carried out, under good seeing ($\sim1.0''$). The data cover three spectral ranges,  [4500 $\mathrm{\AA}$ -- 5450 $\mathrm{\AA}$], [5400 $\mathrm{\AA}$ -- 6750 $\mathrm{\AA}$], and [6700 $\mathrm{\AA}$ -- 7850 $\mathrm{\AA}$]. The RSS long-slit mode was used with the same narrow slit as above resulting in a resolution of $R \sim 7000$ for the first spectral range while a different grating was used for the second and third spectral ranges at a resolution of $R \sim 5000$, with exposure times of  (5 $\times$ 200 s), (4 $\times$ 100 s), and (4 $\times$ 100 s), respectively. The spectra are reduced and calibrated using the  PySALT pipeline \citep{Crawford_etal_2010}.\ The images are combined, the background is subtracted, and the spectra are extracted using the IRAF (Image Reduction and Analysis Facility) software \citep{Tody_1986}.\\
\subsection{Spectral Changes}
\begin{figure*}
\begin{center}
\includegraphics[width=140mm]{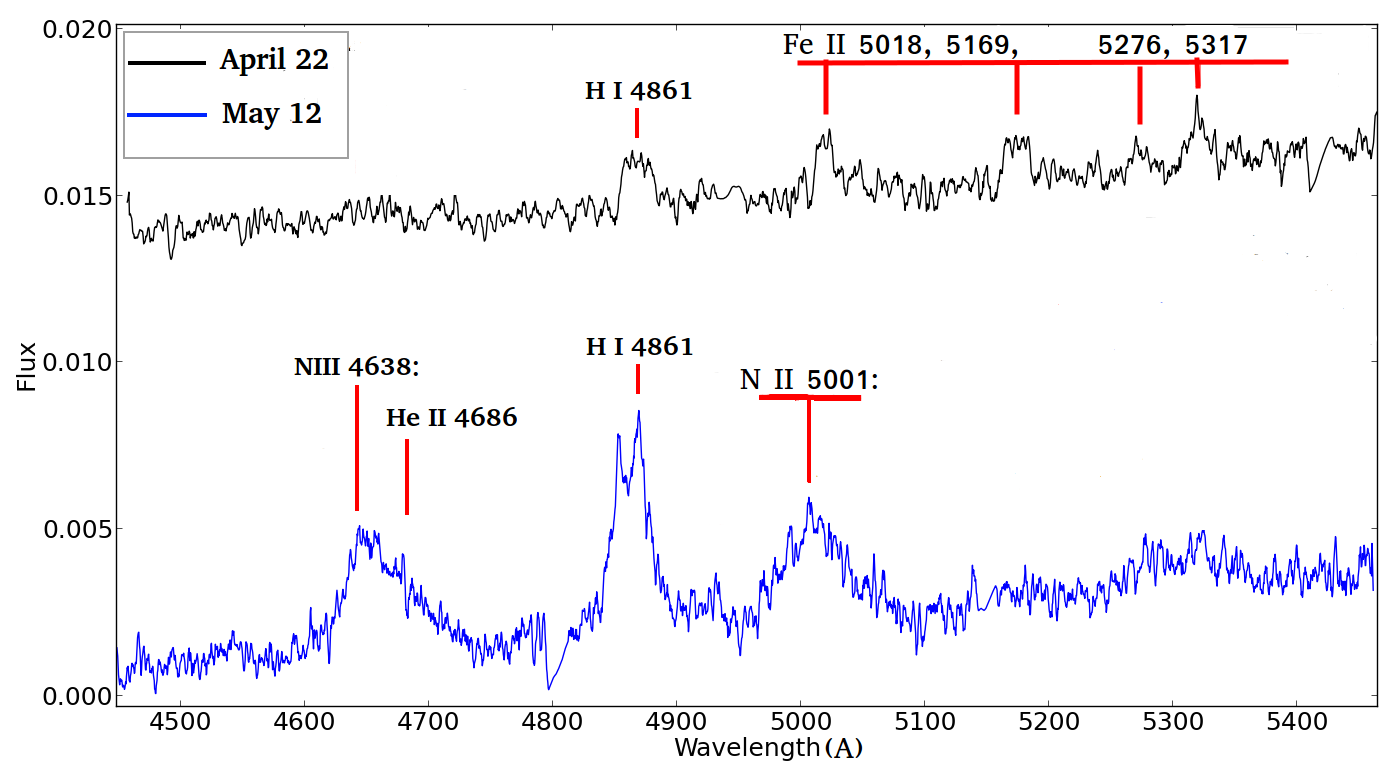}
\caption{The flux plotted to arbitrary units between 4440 $\mathrm{\AA}$ and 5450 $\mathrm{\AA}$. The April 22 spectrum (upper) shows several broad \eal{Fe}{II} emission lines and H${\beta}$ line. However, in the May 12 spectrum (lower), the Fe II lines become weaker and almost disappear in contrast to the He and N that start to appear. The H${\beta}$ line shows a double-peak. For clarity, the April 22 spectrum is vertically shifted. In the April 22 spectrum, chip gaps are between: 4927.2 $\mathrm{\AA}$ and 4953.4 $\mathrm{\AA}$ and between: 5409.0 $\mathrm{\AA}$ and 5433.9 $\mathrm{\AA}$. In the May 12 spectrum, the chip gaps are between: 4790.9 $\mathrm{\AA}$ and 4810.0 $\mathrm{\AA}$ and between: 5140.5 $\mathrm{\AA}$ and 5158.4 $\mathrm{\AA}$.}
\label{Fig:2}
\end{center}
\end{figure*}

\begin{figure*}
\begin{center}
\includegraphics[width=140mm]{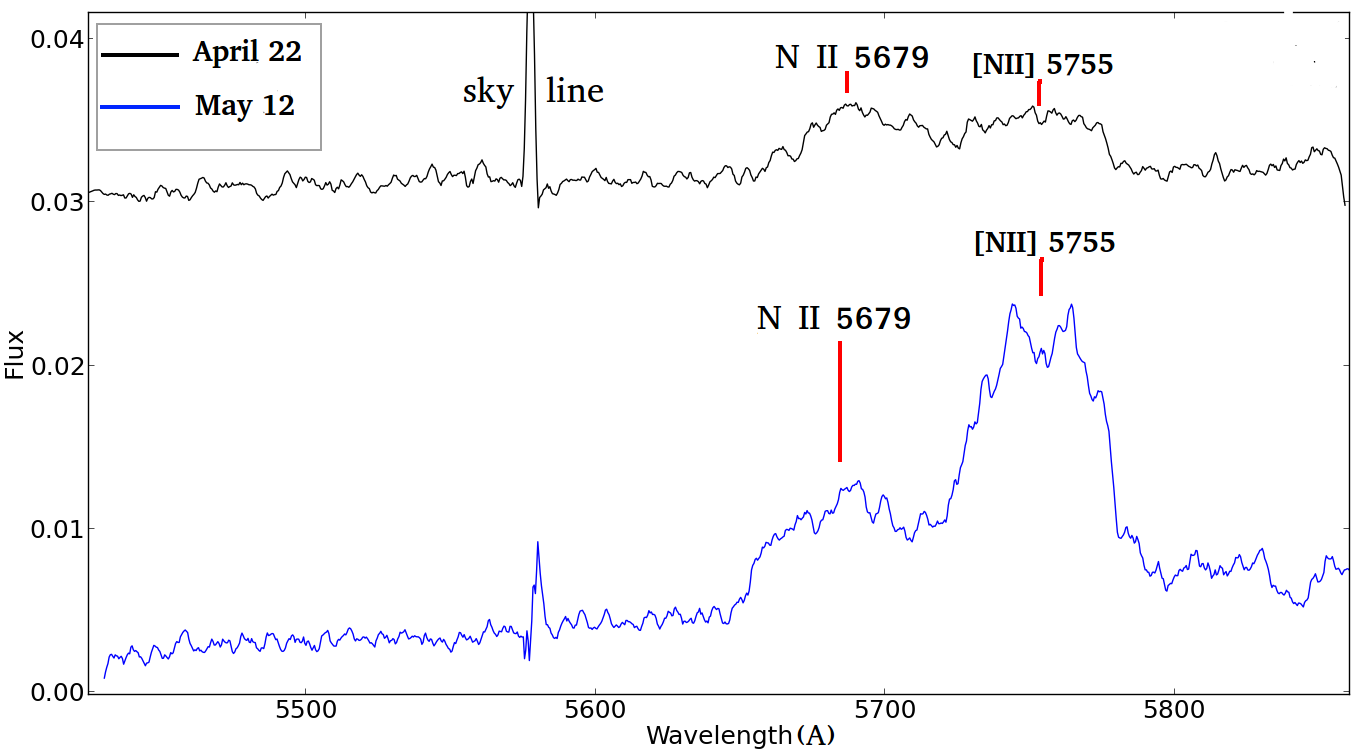}
\caption{As in Fig. \ref{Fig:2}, but} between 5400 $\mathrm{\AA}$ and 5850 $\mathrm{\AA}$. The April 22 spectrum (upper) shows broad and relatively weak \eal{N}{II} 5679 $\mathrm{\AA}$ line and \feal{N}{II} 5755 $\mathrm{\AA}$ line. In the May 12 spectrum (lower), the  \feal{N}{II}  line becomes stronger compared to the N II line. For clarity, the April 22 spectrum is vertically shifted.
\label{Fig:3}
\end{center}
\end{figure*}

\begin{figure*}
\begin{center}
\includegraphics[width=140mm]{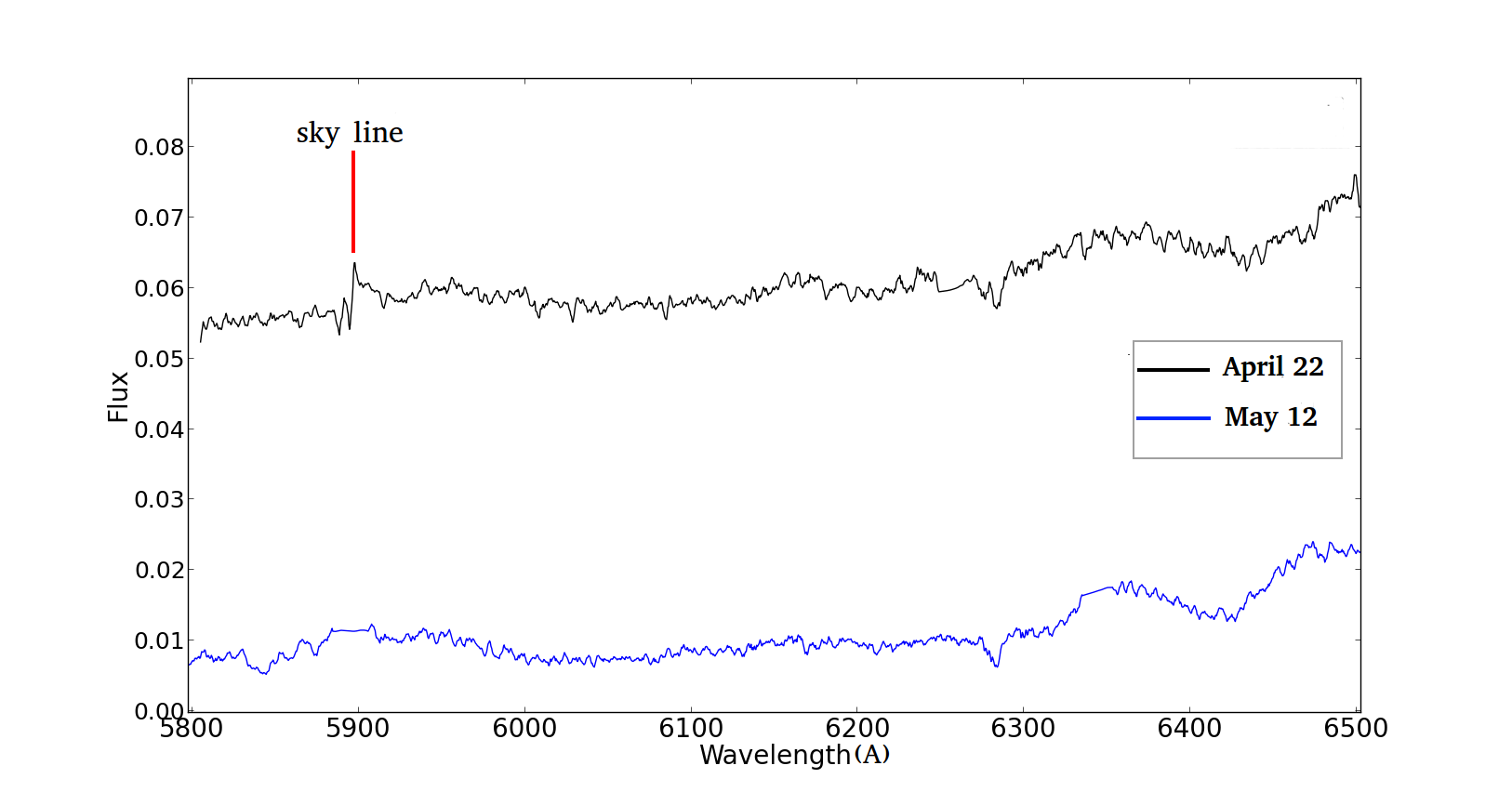}
\caption{As in Fig. \ref{Fig:2}, but between 5850 $\mathrm{\AA}$ and 6500 $\mathrm{\AA}$. No strong emission lines are present in this spectral range. For clarity, the April 22 spectrum is vertically shifted. In the April 22 spectrum, the chip gap is between: 6242.8 $\mathrm{\AA}$ and 6267.1 $\mathrm{\AA}$. In the May 12 spectrum, the chip gaps are between: 5877.9 $\mathrm{\AA}$ and 5902.7 $\mathrm{\AA}$ and between: 6332.6 $\mathrm{\AA}$ and 6355.9 $\mathrm{\AA}$.}
\label{Fig:4}
\end{center}
\end{figure*}

\begin{figure*}
\centering
\begin{subfigure}[b]{0.4\textwidth}
  \centering
 \includegraphics[width=1.5\textwidth]{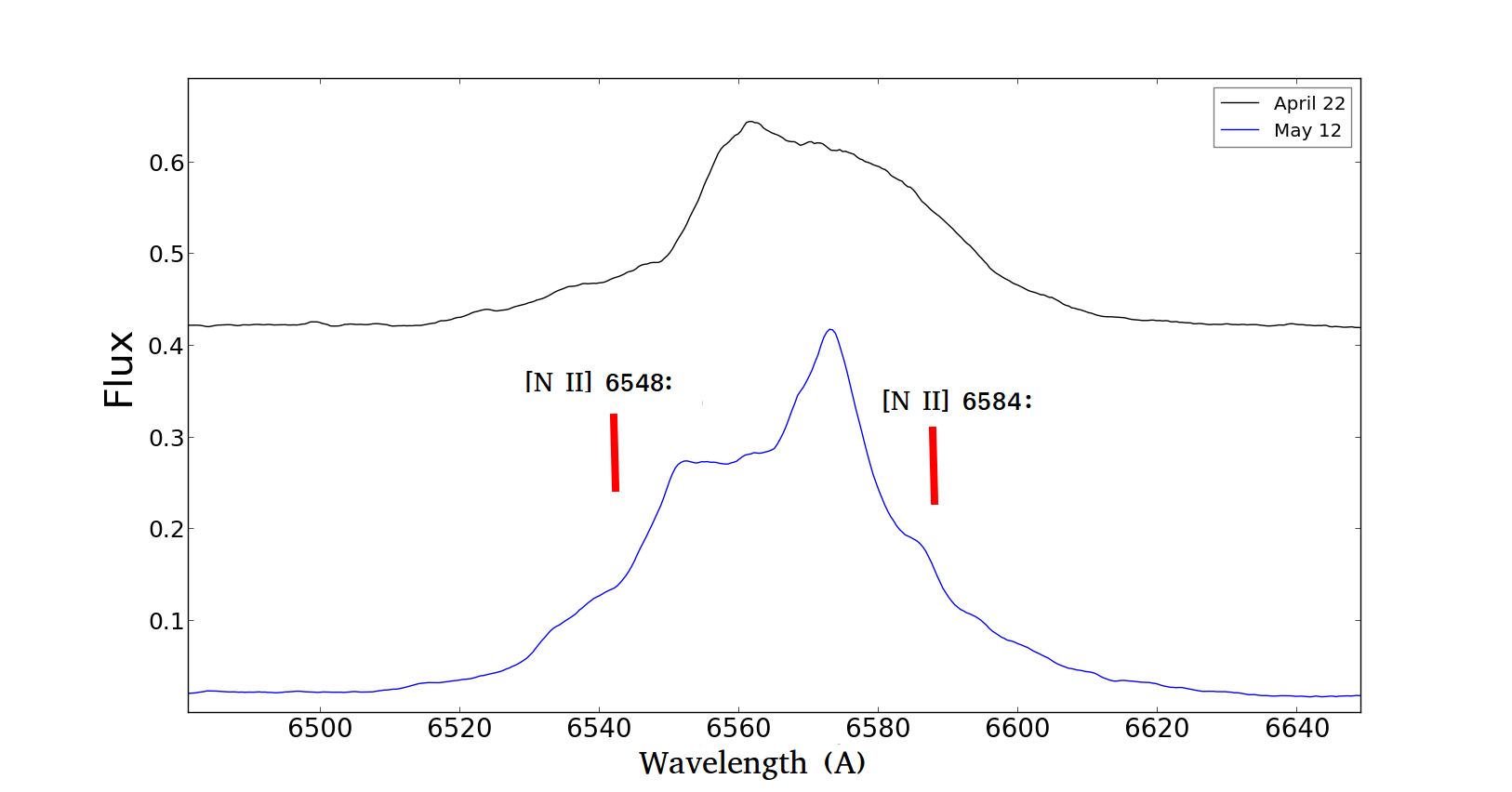}
  \label{fig:5sub1}
\end{subfigure}
\hfill
\begin{subfigure}[b]{0.4\textwidth}
  \centering
 \includegraphics[width=\textwidth]{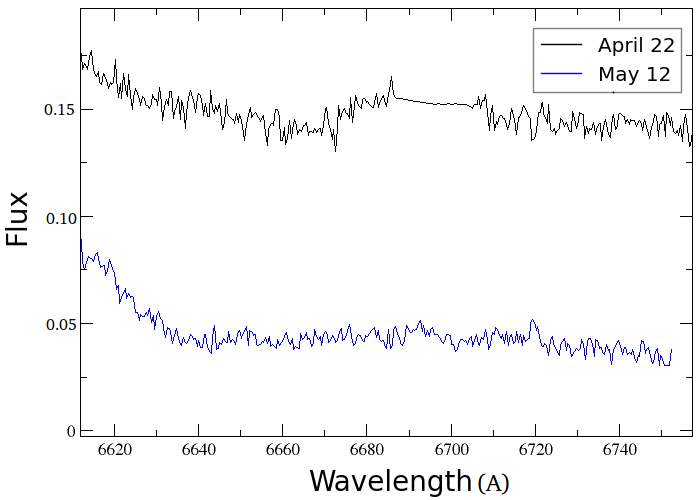}
  \label{fig:5sub2}
\end{subfigure}
\caption{As in Fig. \ref{Fig:2}, but between (a)  6480 $\mathrm{\AA}$ and 6640 $\mathrm{\AA}$, (b) 6620$\mathrm{\AA}$ and 6750 $\mathrm{\AA}$. The April 22 spectrum (upper) shows broad  H${\alpha}$ emission line (FWHM $\sim$ 2300 km s$^{-1}$ $\pm$ 200 km s$^{-1}$). In the May 12 spectrum (lower), the H${\alpha}$ line shows a double-peak similar to the one of H${\beta}$. Both \feal{N}{II} 6548 $\mathrm{\AA}$ and \feal{N}{II} 6584 $\mathrm{\AA}$ are expected to be present but merged with the broad H${\alpha}$. For clarity, the April 22 spectrum is vertically shifted. In the April 22 spectrum, the chip gap is between: 6685.5 $\mathrm{\AA}$ and 6708.1 $\mathrm{\AA}$.}
\label{Fig:5}
\end{figure*}

\begin{figure*}
\begin{center}
\includegraphics[width=140mm]{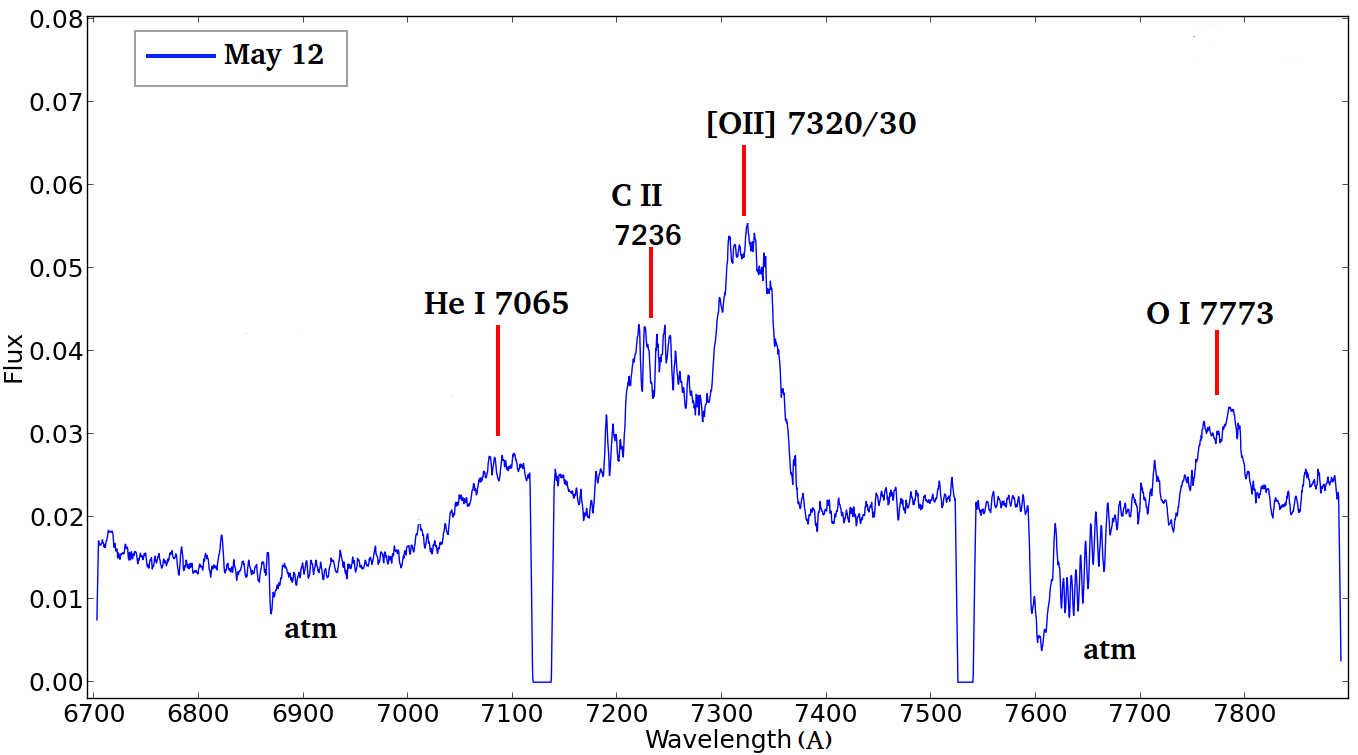}
\caption{As in Fig. \ref{Fig:2}, but between 6750 $\mathrm{\AA}$ and 7850 $\mathrm{\AA}$. The May 12 spectrum shows strong C, He and O emission lines. In the May 12 spectrum, the chip gaps are between: 7111.8 $\mathrm{\AA}$ and 7134.4 $\mathrm{\AA}$ and between: 7521.4 $\mathrm{\AA}$ and 7542.1 $\mathrm{\AA}$.}
\label{Fig:6}
\end{center}
\end{figure*}

In Figs.~\ref{Fig:2} to~\ref{Fig:6}, we show the smoothed spectra, where the top spectrum represents the April 22 observation and the bottom spectrum represents the May 12 observation. The spectral lines were identified mostly using the list from \citet{Williams_2012}. The only clear absorption features in either spectrum are telluric. The April 22 spectra  show several broad flat-topped \eal{Fe}{II}, \eal{H}{I}, and \eal{N}{II} emission lines. However, in the May 12 spectra, the \eal{Fe}{II} lines become weaker and almost disappear, in contrast to the He and N lines that become stronger. H${\beta}$ (Fig.\ref{Fig:2}) and to some extent H${\alpha}$ (Fig.\ref{Fig:5}), show a double peek. These could be the consequence of bipolar ejection or an optically thin shell of gas (see e.g. models of Nova Eri 2009 by \citet{Ribeiro_etal_2013}). Further in the red, between 6750 $\mathrm{\AA}$ and 7850 $\mathrm{\AA}$, the May 12 spectrum shows 
strong He, O and possibly C emission lines.

\subsection{Radial and Expansion Velocities}
\label{specvel}
The Balmer lines  have asymmetric profiles, similar to other novae in the transition stage. In the May spectrum H${\alpha}$ and H${\beta}$ have double peaks with the red peak stronger than the blue one; \feal{N}{II} 5755 $\mathrm{\AA}$, \eal{O}{I} 7773 $\mathrm{\AA}$ and possibly \eal{C}{II} 7236 $\mathrm{\AA}$ are also double peaked.
We derived, using a Lorentzian fitting, the FWHM of H${\alpha}$, H${\beta}$, and \feal{N}{II} 5755 $\mathrm{\AA}$ lines. We found that for H${\alpha}$ the FWHM\,$\sim$\,2300\,$\pm$\,200 km\,s$^{-1}$, for H${\beta}$ the FWHM\,$\sim$\,2400\,$\pm$\,200\,km\,s$^{-1}$, and for \feal{N}{II} 5755 $\mathrm{\AA}$ the FWHM\,$\sim$\,2700\,$\pm$\,200\,km\,s$^{-1}$.

In order to derive the radial velocity of the nova, we measured the H${\alpha}$, H${\beta}$, \eal{Fe}{II} 5018 $\mathrm{\AA}$,  \eal{Fe}{II} 5317 $\mathrm{\AA}$, \eal{N}{II} 5679 $\mathrm{\AA}$, and \feal{N}{II} 5755 $\mathrm{\AA}$ emission lines for the April 22 observation. We also measured the H${\alpha}$, H${\beta}$, \eal{N}{II} 5679 $\mathrm{\AA}$, \feal{N}{II} 5755 $\mathrm{\AA}$, \eal{C}{II} 7326 $\mathrm{\AA}$ and \eal{O}{I} 7773 $\mathrm{\AA}$ emission lines from the second spectrum (May 12). 
For the radial velocity calculations, the rest wavelengths of the emission lines are derived from the multiplet table of astrophysical interest \citep{Moore_1945} and the results listed in Table~\ref{radvel}. In view of the change in the mean velocities of the lines we checked the wavelengths of the overlapping night sky lines with care. We used the European Southern Observatory UVES sky emission spectrum \citep{Hanuschik_2003} for this purpose. The lines at 4861.5 $\mathrm{\AA}$, 5577.5 $\mathrm{\AA}$, 5890 $\mathrm{\AA}$, 5896 $\mathrm{\AA}$, 6300 $\mathrm{\AA}$, 6562 $\mathrm{\AA}$, 6864 $\mathrm{\AA}$, 7276.3 $\mathrm{\AA}$, 7316.1 $\mathrm{\AA}$, 7341 $\mathrm{\AA}$, ... all agree well with the UVES sky emission lines.

 Measurements of permitted lines are potentially shifted to the red by optical depth effects and must be regarded as unreliable. The most reliable measures of a nova radial velocity come from the CNO recombination lines and forbidden lines, which are almost always optically thin. The \feal{O}{II} 7319,7330 $\mathrm{\AA}$ is a blend, so we use only \eal{N}{II} 5679 $\mathrm{\AA}$, \feal{N}{II} 5755 $\mathrm{\AA}$, and \eal{C}{II} 7236 $\mathrm{\AA}$ from the May spectrum to derive the radial velocity. These lines give a mean heliocentric radial velocity of \,$\sim$\,110\,$\pm$\,50\,km\,s$^{-1}$.

The determination of novae radial velocities is not straight forward since many factors can affect it, including the WD orbital motion, optical depth effects, and most importantly the asymmetry of the ejecta \citep{Williams_Bob_1994}. In both observations, the lines are shifted towards the red which would not be expected from optically thick ejecta. 

It is not entirely clear why the radial velocity changes between the two epochs. Although both spectra appear to lack P Cygni absorptions it is possible that the first one was not optically thin and that absorption influences the blue side of the lines shifting them to the red, although the size of the shift makes this unlikely.  It is also possible to speculate that the eruption was very asymmetric and that what we see in the first spectrum is a redshifted clump of material, that later slows, although such an explanation seems contrived.  

The velocity of the Sagittarius stream has not been measured in the direction of  the nova, but 
the central part of the Sagittarius Dwarf Spheroidal galaxy has a mean velocity of 140\,km\,s$^{-1}$ with a velocity dispersion of only 10\,km\,s$^{-1}$ \citep{Ibata_etal_1995}. The mean measurement from the more reliable lines in the second spectrum is consistent with this value. On the other hand the Galactic bulge has a high velocity dispersion and it is certainly not possible to use the velocity to rule out membership of the bulge. 

\citet{Williams_Bob_1994} discusses the radial velocities of Galactic novae that  are mostly within the solar circle and in the direction of the bulge. These have systematically large negative velocities and, as Williams points out, this indicates that either  high internal absorption skews their emission lines to bluer velocities or most of the novae are moving out from the Galactic centre. It is much more difficult to skew the lines of a nova to the red, so if our nova is in the plane, behind the bulge, it would be moving away from the Galactic centre. That possibility cannot be ruled out, but it seems more likely that it is in the bulge, or the Sagittarius stream.


\begin{table}
\begin{center}
\caption{Heliocentric radial velocities of the emission lines for the April 22 and May 12 observations. The uncertainty for all the values is $\pm$ 50 km\,s$^{-1}$.}
\label{radvel}
\begin{tabular}{ccc}
\hline
Line  & $V_{rad}$  (April 22)&  $V_{rad}$ (May 12)  \\
&\multicolumn{2}{c}{$\rm (km\,s^{-1}$)}\\
\hline
H${\alpha}$ & 300 & 155\\
H${\beta}$ & 320 & 140\\
\eal{Fe}{II} 5018 $\mathrm{\AA}$ & 260 & -- \\
\eal{Fe}{II} 5317 $\mathrm{\AA}$ & 180 & --\\
\eal{N}{II} 5679 $\mathrm{\AA}$ & 380  & 140\\
\feal{N}{II} 5755 $\mathrm{\AA}$ & 150  & 45\\
\eal{C}{II} 7236 $\mathrm{\AA}$ & -- & 140\\
\eal{O}{I} 7773 $\mathrm{\AA}$ & -- & 160\\
\hline
\end{tabular}
\end{center}
\end{table}
\subsection{Spectroscopic Classification}

The spectra of V5852 Sgr leave no doubt that this is a CN. CNe are divided into two spectroscopic classes (\eal{Fe}{II} and He/N). The two novae spectral classes have distinctly different properties. He/N novae exhibit broad, high ionization lines, rectangular profiles, few absorption features if any, and a rapidly decreasing visible luminosity. The line broadening is due to high expansion velocities. The spectra are dominated by He, N, and H emission lines. The prominent He and N transitions, the line widths and the rectangular line profiles, are all attributed  to  emission from high velocity gas (WD ejecta) \citep{Williams_2012}.

\eal{Fe}{II} novae have spectra characterized by low excitation narrow \eal{Fe}{II}, \eal{Na}{I} lines, and CNO lines in the far red. They also show prominent P Cygni absorption features characteristic of an optically thick expanding gas. The class of \eal{Fe}{II} can be divided into two sub-classes \eal{Fe}{II}n (narrow) and \eal{Fe}{II}b (broad) where respectively narrow or broad emission lines are present in the spectrum. 
Although no P Cygni absorption features are seen in either spectra, the April 22 spectrum presents features and lines that characterize the \eal{Fe}{II} spectroscopic class. A moderate FWHM of $\sim$ 2500 km s$^{-1}$ and the flat-topped broad lines favors the possibility of a \eal{Fe}{II}b (broad) spectroscopic class. The May 12 spectrum shows several N lines which might indicate a transition into He/N class. However the absence of some helium lines, (\eal{He}{I} 5876 $\mathrm{\AA}$ and 6678 $\mathrm{\AA}$) that are expected to be present for the He/N class suggests that at this stage the nova was actually in transition to the nebular phase.

\section{Discussion $\&$ Conclusions}
\label{Concl}

V5852~Sgr was reported as a possible classical nova by the OGLE sky survey on 2015 March 5. Putting together the nova distance, luminosity, and radial velocity with their uncertainties, leaves the membership of the nova open to three possibilities. The first puts the nova in the Sagittarius stream and the second puts it in the plane behind the bulge moving away from the Galactic centre and the third puts it in  the Galactic bulge (see Section~\ref{specvel}). It is worth noting that the mass of the Sagittarius dwarf  is 10$^8$ -- 10$^9$ M$_{\odot}$ so  the expected nova rate in that galaxy will only be one per 10 to 100 years. The nova rate in the bulge is $\sim$\,2 orders of magnitude higher, so the bulge is a more probable site \citep{Mroz_etal_2015}. So the location will have to remain open, at least until we gain a better understanding of the characteristics of novae with this type of light curve. The membership to the  Sagittarius dwarf galaxy would make it the first nova to be discovered in a dwarf spheroidal galaxy.
\\
The photometric follow up (OGLE, IRSF, LCOGT) revealed a peculiar nova light curve. The spectroscopic and photometric data revealed combined features rarely observed in CNe. The nature of V5852~Sgr is best described as an unusual one.
Based on our spectra and the classification criteria of \citet{Williams_1992} V5852 Sgr shows a \eal{Fe}{II} or \eal{Fe}{II} b spectroscopic class in a transition to the nebular phase.
 Tracking the spectroscopic and photometric evolution of the object is essential for a definitive classification. If the object is a classical, moderately fast nova, the same spectroscopic features should remain until the nebular phase develops. A detailed study of the elemental abundances is also essential. Hence, further observations are needed to definitively identify the object and understand the physical mechanism responsible for this explosion.

\section*{Acknowledgments}

A part of this work is based on observations made with the Southern African Large
Telescope (SALT), under the program 2014-2-DDT-006, and the IRSF, and makes use of observations from the LCOGT network.  The IRSF project is a collaboration between Nagoya University and the South African Astronomical Observatory (SAAO) supported by the
Grants-in-Aid for Scientific Research on Priority Areas (A) (No.  10147207
and No.  10147214) and Optical \& Near-Infrared Astronomy Inter-University
Cooperation Program, from the Ministry of Education, Culture, Sports,
Science and Technology (MEXT) of Japan and the National Research Foundation
(NRF) of South Africa. EA,  PAW, SM, and PV gratefully acknowledge the receipt of research grants from the National Research Foundation (NRF) of South Africa.  We are grateful to Steve Crawford, Marissa Kotze, and Brent Miszalski for assistance with the SALT observations and 
We acknowledge helpful discussions with Joanna Miko\l{}ajewska, Krystian I\l{}kiewicz, Katarzyna Drozd and Massimo Della Valle. We thank Ulisse Munari for his insight on the relative decay rates of novae in the $I$ and $V$ bands and Nikolai  Samus for communicating the variable star name of this nova in advance of publication.\\
P.M. is supported by the ''Diamond Grant'' No. DI2013/014743 funded by the Polish Ministry of Science and Higher Education. We thank the {\it Swift} PI, Neil Gehrels, for an allocation of ToO time. This research was made possible through the use of the AAVSO Photometric All-Sky Survey (APASS), funded by the Robert Martin Ayers Sciences Fund.
The OGLE project has received funding from the National Science Center, Poland, grant MAESTRO 2014/14/A/ST9/00121 to A.U.\\
This work makes use of observations from the LCOGT network, which includes three SUPAscopes owned by the University of St Andrews. The Gaia transient follow-up program uses equal network time allocations from the University of St Andrews and the South African Astronomical Observatory (SAAO).\\

\bibliography{biblio}
\label{lastpage}
\end{document}